\documentclass[prd,showpacs,floatfix,preprint]{revtex4}
\usepackage{epsfig}

\newcommand{\be}{\begin{equation}}
\newcommand{\ee}{\end{equation}}
\newcommand{\bea}{\begin{eqnarray}}
\newcommand{\eea}{\end{eqnarray}}
\newcommand{\DS}{Schwinger--Dyson }
\newcommand{\BS}{Bethe--Salpeter }

\newcommand{\w}{\omega}
\newcommand{\G}{\Gamma}
\newcommand{\s}{\!\cdot\!}

\newcommand{\al}{\alpha}
\newcommand{\ba}{\beta}
\newcommand{\de}{\delta}

\newcommand{\si}{\sigma}

\newcommand{\ga}{\gamma}
\newcommand{\ro}{\rho}
\newcommand{\la}{\lambda}
\newcommand{\ka}{\kappa}
\newcommand{\ta}{\tau}
\newcommand{\et}{\eta}
\newcommand{\ov}[1]{\overline{#1}}
\newcommand{\dk}[1]{\,\,\,\raisebox{-0.4ex}{\large $\bar{}$}\!\!d\,{#1}\,}
\def\slr#1{\setbox0=\hbox{$#1$}           
   \dimen0=\wd0                                 
   \setbox1=\hbox{/} \dimen1=\wd1               
   \ifdim\dimen0>\dimen1                        
      \rlap{\hbox to \dimen0{\hfil/\hfil}}      
      #1                                        
   \else                                        
      \rlap{\hbox to \dimen1{\hfil$#1$\hfil}}   
      /                                         
   \fi}
\begin{document}
\title{Unquenching the Quark-Antiquark Green's Function}
\author{P.~Watson}
\email{peter.watson@theo.physik.uni-giessen.de}
\author{W.~Cassing}
\email{wolfgang.cassing@theo.physik.uni-giessen.de}
\affiliation{Institute for Theoretical Physics, University of Giessen,Heinrich-Buff-Ring 16, 35392 Giessen, Germany}
\begin{abstract}
 We propose a nonperturbative resummation
scheme for the four-point connected quark-antiquark Green's
function $G^4$ that shows how the \BS equation may be `unquenched'
with respect to quark-antiquark loops. This mechanism allows to
dynamically account for hadronic meson decays and
multiquark structures whilst respecting the underlying symmetries.
An initial approximation to the four-point \DS equation --
suitable for phenomenological application -- is examined
numerically in a couple of aspects.  It is demonstrated that this
approximation explicitly maintains the correct asymptotic limits
and contains the physical resonance structures in the near
timelike region in the quark-antiquark channel whereas no
resonances are found in the diquark channel, respectively.
\end{abstract}
\maketitle
\section{Introduction}
For many years, quantum chromo-dynamics (QCD) has been widely
accepted as the underlying theory of the strong interaction.  As
an $SU(N_c)$ gauge-field theory, QCD has as its elementary
particle degrees of freedom the quarks and gluons, whereas the
observables of the strong interaction are the spectra of hadrons
and their spectral properties.  One of the contemporary goals of
QCD is clearly a detailed understanding of how the quark and gluon
fields give rise to the formations of hadrons within a single
theoretical framework. Whereas lattice calculations of QCD have shown
substantial progress in the last years of dealing with dynamical
fermions and improved actions \cite{lattice1}, final and robust
results for light quark masses in the continuum limit are still
lacking nowadays.

In the absence of a final and complete solution to QCD one is motivated to adopt
more pragmatic approaches to gain further insight to the strong interaction and its
bound states.  The \DS equations are the field theoretical
equivalent of the equations of motion for the theory
(see \cite{maris03,roberts00,alkofer00,roberts94} for recent reviews);
they are dynamical equations in the continuum and embody all those symmetries
that define the initial theory such as symmetry under Poincar\'{e} transforms,
the gauge symmetry, the discrete parity and charge conjugation symmetries and
the (broken) chiral symmetry.  The inclusion of all of these vital physical
features gives an impetus to their study.  However, the \DS equations form
an infinite hierarchy of coupled equations that must
be truncated to some degree if actual numerical solutions are addressed.
Contemporary truncation schemes have two
guiding principles: \textit{i}) the explicit maintaining of as many of the
underlying symmetries as feasible and \textit{ii}) agreement of the final
results with observation (and possibly other approaches).

The \DS equations are coupled non-linear integral equations relating the
Green's functions (as `building blocks') of the theory to one another.
By themselves these Green's functions do not have a physical interpretation
but must be combined in various ways to construct physically observable quantities.
One particularly efficacious framework \cite{munczek83,jain93,burden96}
is the \DS equation for the quark propagator
(gap equation) within the rainbow truncation as input into the ladder
\BS equation. The latter is itself a special case of a \DS equation which
leads to a description
of the pion as an (almost massless) Goldstone boson -- associated with chiral
symmetry breaking -- as well as a bound state of two (massive) constituent
quarks. The mass of the constituent quarks here
 is dynamically generated from the spontaneously broken chiral symmetry.
The success of this construction in simultaneously describing two
of the most fundamental aspects of hadron phenomenology can be
traced back to the fact that the truncations employed explicitly
observe the relationship imposed by the (flavor non-singlet) axialvector
Ward-Takahashi identity (AXWTI) 
\cite{maris97b,delbourgo79}.  The AXWTI is an expression for gauge
invariance when applied to quark-axialvector vector Green's
functions. This shows the supreme role of symmetries when applied
to dynamical systems.  The light pseudoscalar and vector meson
masses and leptonic decay constants \cite{maris99a,me02} as well
as electromagnetic form factors \cite{maris02,maris99b} are well
reproduced with few parameters.  In principle such parameters may
be fixed by comparison with related quantities from
lattice-QCD \cite{bhagwat03}.

Whilst the success of the coupled rainbow-ladder \DS-- \BS
framework is laudable, it has so far proved difficult to go
beyond this initial (simplest) truncation scheme. There have been
several attempts to address different aspects of possible
extensions. For example by using effective interactions
characterized by $\de-$functions to resumm classes of diagrams
\cite{bender96,detmold02} and employing hybrid combinations of
finite width and $\de-$function interactions \cite{ongoing04}.
Ironically, it is the AXWTI that  -- preserving the symmetry --
leads to a rapid increase of effort as one increases the level of
sophistication in the truncation scheme.

There are several issues of hadronic physics that the coupled \DS-- \BS framework
has not  addressed so far.  The first of these is the hadronic decay of mesons, e.g.
the decay $\rho \rightarrow \pi \pi$.  This question
has been investigated within the impulse approximation and acceptable results have been
obtained for the decays of vector mesons \cite{jarecke02}; however, the desired
genuinely dynamic description is lacking.  A second issue is the light scalar
spectrum (cf. ref.  \cite{pennington03} for an introduction to this
 topic).  It is not clear presently, whether the lightest scalar mesons
are simple quark-antiquark resonances or if they are dominated by meson-meson, perhaps
even diquark-diquark correlations.  These issues are connected to unquenching, i.e.
the inclusion of internal quark-loops to \BS amplitudes.  The inclusion of such loops
allows for multiple quark-antiquark or quark-quark correlations within the overall
\BS amplitude; these correlations give rise to the decay mechanisms and
internal multiquark structures.  Clearly then, such internal correlations play a
significant role in the dynamical description of hadrons.  The four-point connected
quark-antiquark Green's function ($G^4$) is the simplest (and key quantity) of 
such correlations.

The aim of this paper is: \textit{i}) to motivate the study of the 4-pt quark-antiquark
Green's function with a practical example for future more extended studies, \textit{ii}) 
to show how a phenomenologically useful, dynamically generated approximation can be 
constructed, and \textit{iii}) to show actual numerical results.  The example chosen is the 
unquenching of the \BS kernel in a manner that preserves the AXWTI, maintains 
the charge conjugation properties and allows for the dynamical description of
meson decay widths and possible multiquark systems.

Throughout this paper we work in the isospin ($d=u$) limit with the only distinction
between quark flavors being their current mass.  We take two light flavors of quark,
\textit{up} and \textit{strange}, and consider only the flavor non-singlet eigenstates 
$\ov{u}u$, $\ov{s}u$ and $\ov{s}s$.  This means that we do not consider the effects 
of (isosinglet and singlet) flavor mixing, such as the $U(1)_A$ anomaly induced in 
the $\et-\et'$ system, but rather focus on pure flavor eigenstates.  
We work in Euclidean space throughout, metric $\de_{\mu\nu}$ and with Hermitian 
Dirac matrices that obey $\{\ga_{\mu},\ga_{\nu}\}=2\de_{\mu\nu}$.  
The integral measure is $\dk{k}=d^4k/(2\pi)^4$.

\section{Unquenching the \BS Kernel}

The `unquenching' of a system entails allowing insertion of arbitrary numbers of
internal quark loops into the various amplitudes of the system.
Fully unquenching the system is clearly tantamount to solving a major component
of the theory, which of course is beyond present techniques.  However, as a first
step we consider a restricted class of unquenching terms that include only a
single quark loop and are diagrammatically planar.  Hereafter, we refer to unquenching
as the insertion of this single quark loop.

\begin{figure}[t]
\mbox{\epsfig{figure=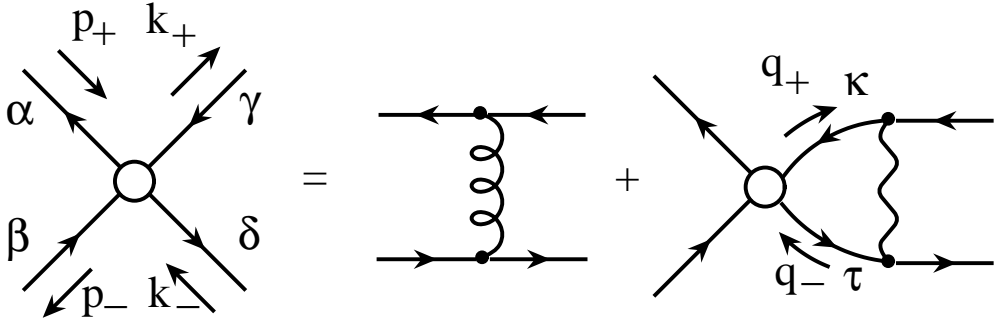,width=14cm}}
\caption{\label{fig:4pt1}\DS equation for the fully amputated, connected quark-antiquark
(4-pt) Green's function, two-particle irreducible with respect to quark-antiquark pairs
under the ladder truncation.  Internal propagators here are fully dressed.}
\end{figure}

As a prelude to unquenching the system we consider the fully amputated, connected
quark-antiquark (4-pt) Green's function $G^4$ that is two-particle irreducible
with respect to any quark-antiquark pair  under the ladder truncation.  It
obeys the \DS equation (shown graphically in Fig.~\ref{fig:4pt1})
\bea
\lefteqn{G_{\al\ba;\de\ga}^4(p_+,p_-;k_-,k_+)=D_{\mu\nu}^{ab}(p_+-k_+)[V_{\mu}^a]_{\al\ga}
[V_{\nu}^b]_{\de\ba}}&&\nonumber\\&&
+\int\dk{q}G_{\al\ba;\ta\ka}^4(p_+,p_-;q_-,q_+)[S_1(q_+)V_{\mu}^a]_{\ka\ga}
[V_{\nu}^bS_2(q_-)]_{\de\ta}\ov{D}_{\mu\nu}^{ab}(q_+-k_+).
\label{eq:4pt1}
\eea
\noindent In the above equation
\be
S_i(p)=-\imath\slr{p}v_i(p^2)+s_i(p^2)
\ee
\noindent denotes the quark propagator of flavor $i$ and
\be
V_{\mu}^a=\imath T^a\ga_{\mu}
\ee
\noindent is the tree-level quark-gluon vertex.  $D_{\mu\nu}^{ab}(q)$ and $\ov{D}_{\mu\nu}^{ab}(q)$
are two different effective interaction terms, one for the seed term and one for the kernel
of the equation. The reader might be worried about two different interactions terms, however,
the reason for this separation will become clear in later sections and allow for practical
truncation schemes that obey the underlying symmetries.
Note, furthermore,  that the distinction of seed and kernel interactions leads to an asymmetry between
left and right quark-antiquark pairs.  The momenta are given by $p_\pm=p\pm Q/2$, similarly for $k$
and $q$.

Equation (\ref{eq:4pt1}) -- under the assumption that the $G^4$ contains resonant
components in the timelike $s$-channel -- can be used as a starting point for the
derivation of the ladder truncation of the \BS equation (see for example
\cite{nakanishi69}).  As mentioned in the
introduction earlier, these resonant components well describe the light pseudoscalar
and vector mesons (the latter albeit in the case where meson decay is neglected).
We emphasize though that Eq. (\ref{eq:4pt1}), unlike the \BS equation, contains both the
resonant \textit{and} non-resonant components in a single expression.

\begin{figure}[t]
\mbox{\epsfig{figure=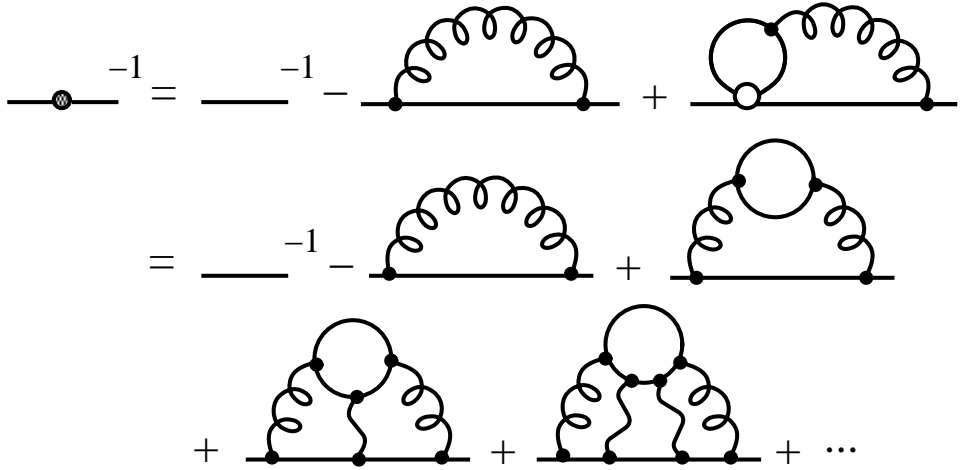,width=16cm}}
\caption{\label{fig:qdse1}Proposed truncation of the quark \DS equation to include
a single quark loop with the use of the 4-pt function $G^4$.  All internal
propagators are dressed.  Springs represent the effective interaction $D$ whereas
wavy lines represent the interaction $\ov{D}$ (see text).}
\end{figure}

With the above 4-pt function ($G^4$) in mind we  propose the truncated quark \DS equation
\bea
\lefteqn{S^{-1}_{\al\ro}(p)=\imath\slr{p}+m-\int\dk{k}[V_{\mu}^aS(k)V_{\nu}^b]_{\al\ro}D_{\mu\nu}^{ab}(p-k)}\nonumber\\&&
+\int\dk{k}\dk{q}G_{\al\ba,\de\ga}^4(p,p+q;k+q,k)[S(p+q)V_{\mu}^aS(k+q)]_{\ba\de}[S(k)V_{\nu}^b]_{\ga\ro}D_{\mu\nu}^{ab}(p-k).\nonumber\\
\label{eq:qdse2}
\eea
\noindent This equation is shown diagrammatically in Fig.~\ref{fig:qdse1}.  The first term
of the self-energy is the standard rainbow truncation.  The next term gives rise to
 unquenching since the 4-pt function $G^4$ (top line of Fig.~\ref{fig:qdse1}), when expanded (lower lines),
generates planar graphs involving a single internal quark loop connected to the original
quark line by successive numbers of gluon exchanges.  By identification of the internal
ladder graphs we collect those terms whose sum has been shown to give rise
 to physical resonances (at timelike momenta).  The exterior interaction factors are
given by $D$, whereas the purely interior interactions are provided by $\ov{D}$; this will
preserve the charge conjugation properties of the quark propagator
(the left-right asymmetry of $G^4$ has been chosen accordingly).
In the unexpanded form of the unquenching term, the explicit occurrence of the interaction
$D$ and the tree-level quark-gluon vertex $V_{\nu}^b$ 
is to maintain the proper counting of the various
graphs in the expansion.  The explicit factoring of one tree-level vertex in each dressing
term is a general feature of \DS equations and is maintained here.  Equations (\ref{eq:qdse2})
and (\ref{eq:4pt1}), once $D$ and $\ov{D}$ have been specified, form a closed system which
can be solved numerically and naturally includes possible resonance structures in
the integrand at timelike momenta.  One can regard the multiple gluon exchange 
terms in the unquenched quark \DS equation as either a correction to the gluon 
propagator and/or vertex function or as a nonperturbative dressing of the quark by 
higher order correlations.

The homogeneous quark-antiquark \BS equation is
\be
\G_{\al\ba}(p;P)=\int\dk{k}K_{\al\ba;\de\ga}(p,k;P)\left[S_1(k_+)\G(k;P)S_2(k_-)
\right]_{\ga\de}
\label{eq:bse}
\ee
\noindent and the pole condition $P^2=-M^2$ can be found once the kernel $K$ and the quark
propagators $S_i$ are specified.  As emphasized earlier,  the pion can be
interpreted as the Goldstone boson of chiral symmetry breaking \cite{maris97b}
when the kernel and the quark self-energy are related such that the AXWTI is satisfied.
The flavor non-singlet AXWTI can be expressed in the following way \cite{ongoing04}
\be
[S^{-1}(p_+)\ga_5+\ga_5S^{-1}(p_-)]_{\al\beta}=[-\imath\ga_5\slr{P}+2m\ga_5]_{\al\beta}+\int\dk{k}K(p,k;P)_{\al\beta,\de\ga}[\ga_5S(k_-)+S(k_+)\ga_5]_{\ga\de}.
\label{eq:axwti1}
\ee
\noindent There are two facets to this relation: on one hand, as a Ward-Takahashi type
identity it is the expression of the underlying gauge symmetry written in a
particular channel and is genuinely nonperturbative in nature; on the other
hand it can be expanded semi-perturbatively to reveal relationships
between different types of graphs.  The former interpretation is responsible
for the chiral symmetry considerations and for the hope that -- whilst one may
not be solving the entire theory --  one might be able to realistically describe
a variety of physical phenomena.  The latter interpretation allows  to develop
kernels $K$ even though the relation introduced above is an integral relationship
involving nonperturbative objects.

\begin{figure}[t]
\mbox{\epsfig{figure=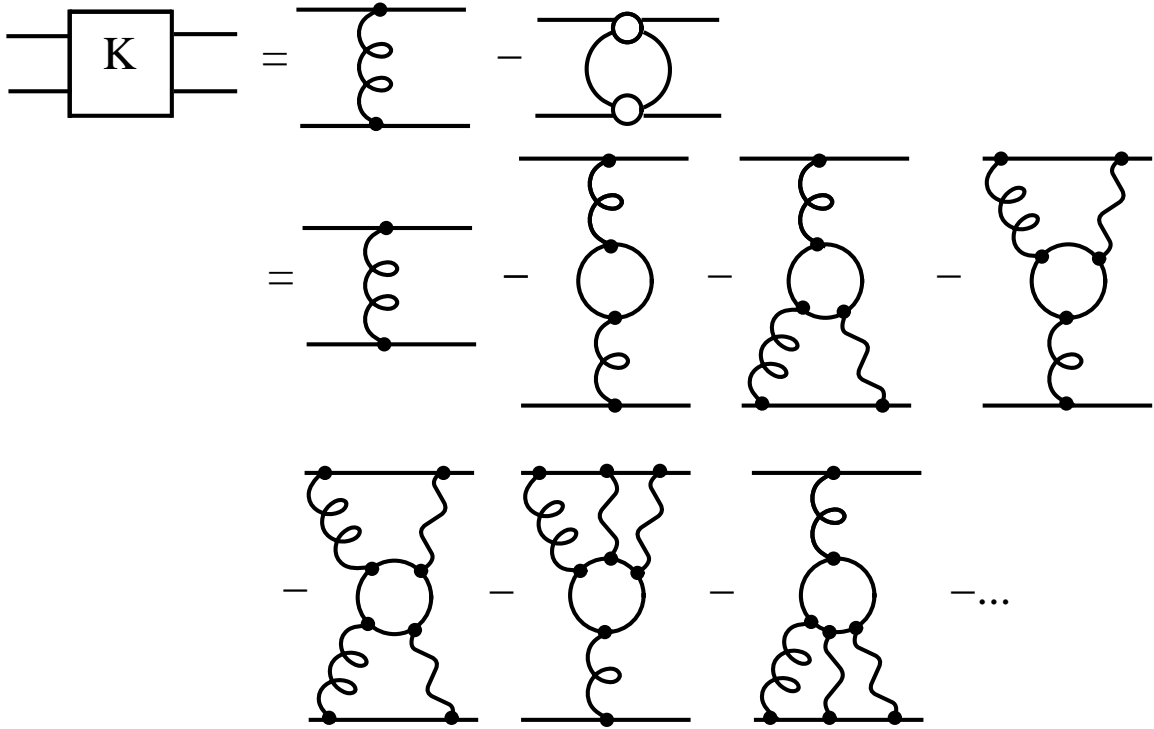,width=16cm}}
\caption{\label{fig:kern1}Proposed truncation of the \BS kernel to include a single
quark loop.  All internal propagators are dressed.  Springs represent the effective
interaction $D$ whereas wavy lines represent the interaction $\ov{D}$ (see text).}
\end{figure}

We first introduce the kernel and show how it and the quark propagator from Eq. (\ref{eq:qdse2})
satisfy the AXWTI, Eq. (\ref{eq:axwti1}). A short discussion of the physical
implications will be given later on.  The kernel is
\bea
\lefteqn{K_{\al\ba;\de\ga}(p,k,P)=D_{\mu\nu}^{ab}(p-k)[V_{\mu}^a]_{\al\ga}[V_{\nu}^b]_{\de\ba}}\nonumber\\&&
-\int\dk{q}G^4_{\al\et;\si\ga}(p_+,p+q;k+q,k_+)G^4_{\ta\ba;\de\ka}(p+q,p_-;k_-,k+q)[S(p+q)]_{\et\ta}[S(k+q)]_{\ka\si},\nonumber\\
\label{eq:kern1}
\eea
\noindent shown diagrammatically in Fig.~\ref{fig:kern1}.  Just as for the quark self-energy
this kernel contains the tree-level (ladder) term and  a contribution involving
$G^4$.  The expansion of $G^4$ into its diagrammatic form  shows how a single
quark loop is incorporated and each graph is planar.  The kernel clearly has the correct
charge conjugation properties.  Knowing that the tree-level, rainbow quark and ladder \BS terms
already satisfy the AXWTI, Eq. (\ref{eq:axwti1}), we eliminate these terms.  Using the
diagrammatic expansion of the kernel and following the quark line, one sees that
starting from left (index $\al$) to right (index $\ba$) or vice-versa on either
side of Eq. (\ref{eq:axwti1}) the quark first and also finally interacts with the effective
interaction $D$; all interactions between (if any) are with $\ov{D}$.  The quark
loop, i.e. the only object proportional to the number of quark flavors $N_f$, is left
untouched on both sides of the equation.  This ordering and the fact that all planar
diagrams are included imply that the AXWTI is naturally satisfied.  This can be
straightforwardly verified explicitly to arbitrary order without explicit specification
of $D$ or $\ov{D}$.

Having established that the charge conjugation properties of and the AXWTI connecting
the quark \DS equation and \BS kernel are satisfied, we now turn to the interpretation
of the unquenching terms within the \BS equation.  There is the dual description of $G^4$
as: \textit{i}) a sum of ladder diagrams and \textit{ii}) as a quark-antiquark correlation function.  Recalling
that the ladder approximated $G^4$ contains good representations of the physical pions
and kaons, the integral term of the kernel will now involve imaginary components picked
up from integration over these resonances.  This in turn gives the possibility for an
imaginary component to the meson mass solution of the \BS equation reflecting its
finite hadronic decay width.  In this way the dynamical description of meson decays
may be achieved as well as a possible dynamical description of multiquark
states and their mixing with the quark-antiquark components.  The real contributions
from the non-resonant parts of $G^4$ will only modify the position of the overall resonance.

As a further remark, the form of the kernel makes clear that only flavor non-singlet
combinations are included within this unquenching scenario.  In fact, one can add 
more terms to the kernel that account for both the flavor singlet mixing and
the anomalous breaking of chiral symmetry.  Allowing for flavor singlet contributions 
one may `open' the unquenching quark loop to form a kernel with two or more 
gluon exchanges in the s-channel.  A similar system has been considered 
previously \cite{kekez00,kogut74}, though with only two-gluon exchange.  The 
kernel is still consistent with the AXWTI, up to anomalous terms.  The reason 
for this is that while the quark self-energy can be constructed from functional
derivatives of an effective action, the kernel is an additional functional 
derivative of the quark self-energy \cite{maris03}.  The flavor singlet 
contributions are then the functional derivative of the unquenching quark loop in 
our constructed quark self-energy Eq. (\ref{eq:qdse2}).  In this work, however, 
the kernel Eq. (\ref{eq:kern1}) only involves the functional derivatives 
of the original quark line and hence only involves flavor non-singlet parts.

From the separate unquenching diagrams the quark loop is expressed as a trace over
Dirac matrices; separately, each diagram is independent of the direction of the
quark line.  If the sum of these diagrams is replaced by the corresponding full 4-pt
functions then again the direction of the internal quark lines is irrelevant.
However, for
approximate functions this is no longer true.  We choose to use quark-antiquark 4-pt
functions since we expect their resonances to form the basis for the
physical decay of the overall meson.  This outlines an important aspect of unquenching
phenomena. We mention that with quark-antiquark internal 4-pt functions one
naturally expects an
internal structure composed of mesons (and of course the original quarks) but the
full 4-pt function contains also (non-resonant) diquark correlations.
Moreover, one could in principle use quark-quark internal 4-pt functions and a
diquark basis instead.  The goal, therefore, is to maintain a reliable
approximation to $G^4$ -- in both resonant and non-resonant channels -- such that
 the direction of the quark line does not affect results for physical
observables.  This picture of unquenching, where components are dynamically coupled to the
quark-antiquark parts, serves as an illustration of the phenomenological
descriptions of scalar mesons as $q\ov{q}$ \cite{scadron03}, meson-meson
\cite{tornqvist02}, diquark-antidiquark \cite{jaffe77,black00} or combinatoric
\cite{close02} in character. In practice, all models can yield good results; the different descriptions
essentially reproduce different parts of the full dynamical system.

\section{A Tractable Model}

The Eqs. (\ref{eq:4pt1}), (\ref{eq:qdse2}), (\ref{eq:bse}) and (\ref{eq:kern1}) -- once $D$
and $\ov{D}$ have been specified -- can be used \textit{in principle} to find the
resonance mass of a given channel while including a single internal quark loop.
The major hindrance to solving this system is the number of components that must
be kept track of.  We here propose a suitable approximation
that is expected to keep track of the most important pieces,  and omit the rest for
future study.

The first step -- in solving the unquenched \BS system -- is the specification
of the effective interactions $D$ and $\ov{D}$.  With the rainbow-ladder coupled
\DS-- \BS framework (which only uses $D$), the Landau gauge and ultraviolet suppressed
 form
\be
D_{\mu\nu}^{ab}(q)=\de^{ab}t_{\mu\nu}(q)D(q^2)=\de^{ab}t_{\mu\nu}(q)4\pi^2D\frac{q^2}{\w^2}\exp{\left(-\frac{q^2}{\w^2}\right)}
\label{eq:g1}
\ee
\noindent has proven to provide for a good description of the light
pseudoscalar and vector mesons \cite{me02} (in the absence of
hadronic decays).  The form of this effective interaction is
tailored to primarily give results for hadron observables such as
masses and weak decay constants.  The success of the
interaction Eq. (\ref{eq:g1}) can be attributed to the concept of
integrated infrared strength -- the explicit shape of the
parameterization is unimportant compared to the integral area in
the infrared region.  One may add ultraviolet logarithmic terms
to provide the correct perturbative limit at high momenta or 
compare with lattice calculations but these details, though 
important, are not the issue here (see e.g. \cite{maris03} for a more
comprehensive discussion of such issues).  From the construction
of the unquenched system it has become clear that $D$ is the
leading interaction term while $\ov{D}$ is sub-leading. Since the
AXWTI can be satisfied even for a separate specification of the
interaction terms, we have some flexibility in constructing the
complimentary interaction $\ov{D}$.  Our initial criteria for choosing the
form of $\ov{D}$ are that the resonance structure of $G^4$ for the
lightest pseudoscalar mesons should be well approximated and that
Eq. (\ref{eq:4pt1}) is tractable numerically. We choose
\be
\ov{D}_{\mu\nu}^{ab}(q)=\de^{ab}\de_{\mu\nu}\ov{D}(q^2)=\de^{ab}\de_{\mu\nu}\frac{3}{4}D(q^2).
\label{eq:g2}
\ee
\noindent This form is similar to Eq. (\ref{eq:g1}) but is now in Feynman-like
gauge.  As an effective interaction it reproduces the leading structure of the
pseudoscalar rainbow \BS equation within Eq. (\ref{eq:4pt1}) and
retains the desired approximation to the $\pi$ and $K$ mesons.  It
will also be seen that it allows for a Fierz reordering of the
Dirac matrices of Eq. (\ref{eq:4pt1}); in this way the 4-pt quark
problem reduces to a single spin line.

The next approximation is to neglect the back-reaction of the 4-pt function in the
quark self-energy, i.e. we employ only the rainbow quark propagator.  We stress that this
is an initial approximation performed for heuristic purposes; the self-consistent set of
Eqs. (\ref{eq:4pt1}) and (\ref{eq:qdse2}) will have to be examined in future in order to have
a control on the approximations made here.   In mitigation though, we posit that the major
contribution of this
term is to add possible imaginary parts and affect the resonance structure of the quark
propagator functions in the timelike region.  We expect that in the immediate vicinity of the
spacelike axis (the region of interest here) there should be only small effects.

The quark propagator is not itself an observable quantity and the
precise details of its dressing are not directly of concern. One
consequence of quark propagator dressing is the vacuum chiral
quark condensate $<\ov{q}q>_0 \neq 0$, which is a quantitative
measure of dynamical chiral symmetry breaking \cite{tandy97}. The
condensate  $<\ov{q}q>_0$ is calculable as an integral of the
scalar part of the quark propagator at spacelike momenta alone and
is real-valued. This gives credence to the assertion that the
inclusion of the back-reaction term will not qualitatively alter
the picture for spacelike momenta.  Since $<\ov{q}q>_0$ is closely
related to the properties of the pion, any minor quantitative
changes could be compensated for by re-fitting model parameters.
We also note that in \cite{bhagwat02}, where quark propagators
constructed from simple complex conjugate poles have been
investigated, the imaginary parts in the \BS equation -- generated
by integration over these poles -- canceled exactly, i.e. the
kernel forbids meson decay into quark constituents.  Thus,
although the details of the quark propagator in the timelike
region will not be exactly correct when omitting the unquenched
term, we do not expect the overall results to be seriously
deficient.

$G^4$ is decomposed into color components,
\be
G_{\al\ba;\de\ga}^4=\de_{\al\ba}^c\de_{\de\ga}^cG_{\al\ba;\de\ga}^m+\de_{\al\ga}^c\de_{\de\ba}^cG_{\al\ba;\de\ga}^n,
\ee
\noindent where (for a given flavor combination input via the $S_i$) $G^m$
and $G^n$ carry only Dirac and Lorentz structure.  $G^m$
corresponds to a mesonic correlation whereas $G^n$ can be related
to a diquark correlation.  In the following we refer to $G^n$ as
the diquark channel.

Inserting the tree-level vertices, using the color identity \cite{cvitanovic76}
\be
[T^a]_{\al\ga}[T^a]_{\de\ba}=\frac{1}{2}\de_{\al\ba}^c\de_{\de\ga}^c-\frac{1}{6}\de_{\al\ga}^c\de_{\de\ba}^c,
\ee
\noindent the spin Fierz identity \cite{cahill89}
\be
[\ga_{\mu}]_{\al\ga}[\ga_{\mu}]_{\de\ba}=[K^a]_{\al\ba}[K^a]_{\de\ga},\;K^a=\{\openone,\imath\ga_5,\imath\ga_{\la}/2,\imath\ga_5\ga_{\la}/2\}
\ee
\noindent and inserting the effective interaction forms
Eq. (\ref{eq:g1}) and Eq. (\ref{eq:g2}), the expression for the 4-pt function,
Eq. (\ref{eq:4pt1}), becomes
\bea
G_{\al\ba;\de\ga}^n(p_+,p_-;k_-,k_+)\!\!\!\!&&=\frac{D(t)}{6}\left\{[K^a]_{\al\ba}[K^a]_{\de\ga}-\frac{(\slr{p}-\slr{k})_{\al\ga}(\slr{p}-\slr{k})_{\de\ba}}{t}\right\}\nonumber\\&&
+\frac{1}{6}\int\dk{q}\ov{D}((q-k)^2)G_{\al\ba;\ka\et}^n(p_+,p_-;q_-,q_+)\left[S_1(q_+)K^aS_2(q_-)\right]_{\et\ka}[K^a]_{\de\ga}\nonumber\\
G_{\al\ba;\de\ga}^m(p_+,p_-;k_-,k_+)\!\!\!\!&&=-3G_{\al\ba;\de\ga}^n(p_+,p_-;k_-,k_+)\nonumber\\&&
-\frac{4}{3}\int\dk{q}\ov{D}((q-k)^2)G_{\al\ba;\ka\et}^m(p_+,p_-;q_-,q_+)\left[S_1(q_+)K^aS_2(q_-)\right]_{\et\ka}[K^a]_{\de\ga}\nonumber\\
\eea
\noindent where $t=(p-k)^2$.  Rewriting the solutions as
\bea
G_{\al\ba;\de\ga}^n(p_+,p_-;k_-,k_+)&=&G_{\al\ba}^{n,a}(p_+,p_-;k_-,k_+)[K^a]_{\de\ga}-\frac{D(t)}{6t}(\slr{p}-\slr{k})_{\al\ga}(\slr{p}-\slr{k})_{\de\ba}\nonumber\\
G_{\al\ba;\de\ga}^m(p_+,p_-;k_-,k_+)&=&G_{\al\ba}^{m,a}(p_+,p_-;k_-,k_+)[K^a]_{\de\ga}+\frac{D(t)}{2t}(\slr{p}-\slr{k})_{\al\ga}(\slr{p}-\slr{k})_{\de\ba}\nonumber\\
\label{eq:sol}
\eea
\noindent reduces the 4-pt equations to a problem involving a single spin-line
\bea
G_{\al\ba}^{n,a}(p_+,p_-;k_-,k_+)\!\!\!\!&&=\frac{D(t)}{6}[K^a]_{\al\ba}\nonumber\\&&
-\frac{1}{36}\int\dk{q}\ov{D}((q-k)^2)\frac{D((p-q)^2)}{(p-q)^2}\left[(\slr{p}-\slr{q})S_1(q_+)K^aS_2(q_-)(\slr{p}-\slr{q})\right]_{\al\ba}\nonumber\\&&
+\frac{1}{6}\int\dk{q}\ov{D}((q-k)^2)G_{\al\ba}^{n,b}(p_+,p_-;q_-,q_+)Tr\left[K^bS_1(q_+)K^aS_2(q_-)\right]\nonumber\\
G_{\al\ba}^{m,a}(p_+,p_-;k_-,k_+)\!\!\!\!&&=-3G_{\al\ba}^{n,a}(p_+,p_-;k_-,k_+)\nonumber\\&&
-\frac{2}{3}\int\dk{q}\ov{D}((q-k)^2)\frac{D((p-q)^2)}{(p-q)^2}\left[(\slr{p}-\slr{q})S_1(q_+)K^aS_2(q_-)(\slr{p}-\slr{q})\right]_{\al\ba}\nonumber\\&&
-\frac{4}{3}\int\dk{q}\ov{D}((q-k)^2)G_{\al\ba}^{m,b}(p_+,p_-;q_-,q_+)Tr\left[K^bS_1(q_+)K^aS_2(q_-)\right].
\eea
\noindent We stress that it is the choice of Feynman-like gauge
for $\ov{D}$ that enables us to exploit the spin 
Fierz identity such that the Dirac structure of $G^{m,n}$ can 
be reduced with Eq. (\ref{eq:sol}).  This reduction makes the analysis 
of $G^4$ feasible for phenomenological purposes.

To proceed we make two further approximations (again for heuristic
convenience).  The first is to drop any terms in the trace
proportional to the vectors $q_{\pm\la}$, i.e. we restrict to the
leading covariant terms.  This has the effect of suppressing any
mixing between different channels, thus leaving only diagonal
elements. In the spin-1 case only the terms proportional to the
metric then survive.  The second approximation is to drop the
second term on the right-hand side of each equation.  These last
approximations allow us to write the original 4-pt function and
its \DS equation in a much simplified form, which can be
numerically evaluated ($s=(p_+-p_-)^2=Q^2$):
\bea
\lefteqn{G_{\al\ba;\de\ga}^4(p_+,p_-;k_-,k_+)=\frac{D(t)}{t}(\slr{p}-\slr{k})_{\al\ga}(\slr{p}-\slr{k})_{\de\ba}\left[\frac{1}{2}\de_{\al\ba}^c\de_{\de\ga}^c-\frac{1}{6}\de_{\al\ga}^c\de_{\de\ba}^c\right]}\nonumber\\&&
+[K^a]_{\al\ba}[K^a]_{\de\ga}\left[\de_{\al\ba}^c\de_{\de\ga}^cG^{m,a}(s,t;k^2,k\s Q)+\de_{\al\ga}^c\de_{\de\ba}^cG^{n,a}(s,t;k^2,k\s Q)\right]
\eea
\noindent where the (Lorentz scalar) functions $G^{m,a}$ and $G^{n,a}$ are the solutions of
\bea
G^{n,a}(s,t;k^2,k\s Q)&=&\frac{D(t)}{6}+\frac{1}{6}\int\dk{q}\ov{D}((q-k)^2)G^{n,a}(s,t;q^2,q\s Q){\cal K}^a\nonumber\\
G^{m,a}(s,t;k^2,k\s Q)&=&-3G^{n,a}(s,t;k^2,k\s Q)-\frac{4}{3}\int\dk{q}\ov{D}((q-k)^2)G^{m,a}(s,t;q^2,q\s Q){\cal K}^a\nonumber\\
\label{eq:4pt3}
\eea
\noindent with the kernels
\bea
{\cal K}^1&=&4s_1(q_+^2)s_2(q_-^2)-4q_+\s q_-v_1(q_+^2)v_2(q_-^2)\nonumber\\
{\cal K}^5&=&-4s_1(q_+^2)s_2(q_-^2)-4q_+\s q_-v_1(q_+^2)v_2(q_-^2)\nonumber\\
{\cal K}^{\la}&=&-s_1(q_+^2)s_2(q_-^2)-q_+\s q_-v_1(q_+^2)v_2(q_-^2)\nonumber\\
{\cal K}^{5\la}&=&s_1(q_+^2)s_2(q_-^2)-q_+\s q_-v_1(q_+^2)v_2(q_-^2).
\eea

There are several remarks in order here: The seed term in 
Eqs. (\ref{eq:4pt3}) contains the only occurrence of the variable $t$.
Since the solution will be proportional to this term, we can
numerically set it equal to $1/\sqrt{2}$ (see the next paragraph
for an explanation of the Chebyshev convention) and thereafter
express the solution in units of $\sqrt{2}D(t)/6$.  The function
$D$ is  the original effective interaction dressing function and
so the $t$ dependence of $G^4$ is explicitly given by the single
gluon exchange of the original seed term of Eq. (\ref{eq:4pt1}).  We
also  note that as $k^2\rightarrow\infty$, $G^{n,a}(s,t;k^2,k\s
Q)\rightarrow D(t)/6$, $G^{m,a}(s,t;k^2,k\s Q)\rightarrow
-D(t)/2$.  This is seen by noting that $\ov{D}((q-k)^2)$ vanishes
exponentially as $k^2\rightarrow\infty$ unless $q^2\sim
k^2\rightarrow\infty$ in which case ${\cal K}^a$ vanishes.  These
considerations show that due to the form of the 4-pt dressing
function the asymptotic limits of $G^4$ are preserved. Finally,
the \BS equation, albeit in approximated form, is still contained
within Eqs. (\ref{eq:4pt3}) with the position of the resonances
dependent solely on $s$.  This is important since we need to
identify explicitly the $1/(s+m^2)$ factors corresponding to meson
or diquark propagation in order to specify the propagators for
future applications.

In order to solve Eqs. (\ref{eq:4pt3}) for a particular value of $s$,
we perform a Chebyshev expansion in the angular variables $z=k\s
Q/\sqrt{k^2Q^2}$ (and correspondingly $z'=q\s Q/\sqrt{q^2Q^2}$)
such that
\be
G^{(n,m),a}(s,t;k^2,k\s Q)\rightarrow G^{(n,m),a}(s,t;k^2,z)=\sum_{i=0}^{N-1}G_i^{(n,m),a}(s,t;k^2)\,T_i(z).
\ee
\noindent Now the equations are a set of integral equations in a single
variable ($k^2$). Note that we use a convention where
$T_0=1/\sqrt{2}$.  Equations (\ref{eq:4pt3})
are Fredholm equations of the 2nd kind, which can be solved for
arbitrary complex $s$ firstly for spacelike $k^2$ and then for
general complex $k^2$.

\section{Results for the 4-pt Function}

In this section we present numerical results for the components of
the 4-pt function $G^4$.  In what follows we use the parameters
$\w=0.5GeV$, $D=16GeV^{-2}$, $m_u=5MeV$, $m_s=115MeV$ (taken from
\cite{me02}) and we set $N=4$.  The (rainbow truncated) quark
propagator functions are the same as in ref. \cite{me02}.   Looking at the form
of the kernel Eq. (\ref{eq:kern1}) we need to evaluate the functions
$G_i^{(n,m),a}(s,t;l^2)$ at $Q=P/2\mp q$ ($s=Q^2$) and $l=k+q/2\pm
P/4$, where $P^2=-M^2$ is the complex mass solution of the \BS
equation, and $k,q$ are spacelike momentum integration variables.

The positions of the timelike singularities for the mesonic
components $G^m$ (with the above parameters) are given in Table
\ref{tab:sing1}.  The (lightest) scalar and pseudoscalar results
are surprisingly good, given the level of approximations applied.
The vector and axialvector resonance positions are clearly
deficient and this is especially severe in the asymmetric
$\ov{s}u$ flavor case.  However, in our present study of
unquenching effects in the light meson spectrum only the lightest
resonances will be of importance due simply to the kinematics.
\begin{table}[t]
\begin{tabular}{|c||c|c|c||c|c|c|}\hline
     &\multicolumn{3}{|c||}{Feynman-like}&\multicolumn{3}{c|}{\cite{me02}}\\\hline
$J^P$&$\ov{u}u$&$\ov{s}u$&$\ov{s}s$&$\ov{u}u$&$\ov{s}u$&$\ov{s}s$\\\hline\hline
$0^+$&656&957&1230&645&903&1113\\\hline
$0^-$&130&446&610&137&492&---\\\hline
$1^-$&912&1540&1362&758&946&1078\\\hline
$1^+$&1116&$>2000$&1656&915&1085&1233\\\hline
\end{tabular}
\caption{\label{tab:sing1}Masses of mesonic resonances ($MeV$) in the 4-pt function $G^4$.
Results are compared with those of \cite{me02} which were calculated with all covariant terms,
stable with respect to the number of Chebyshev moments (and shown to be explicitly Poincar\'e
covariant).}
\end{table}

\begin{figure}[t]
\vspace{1cm}
\mbox{\epsfig{figure=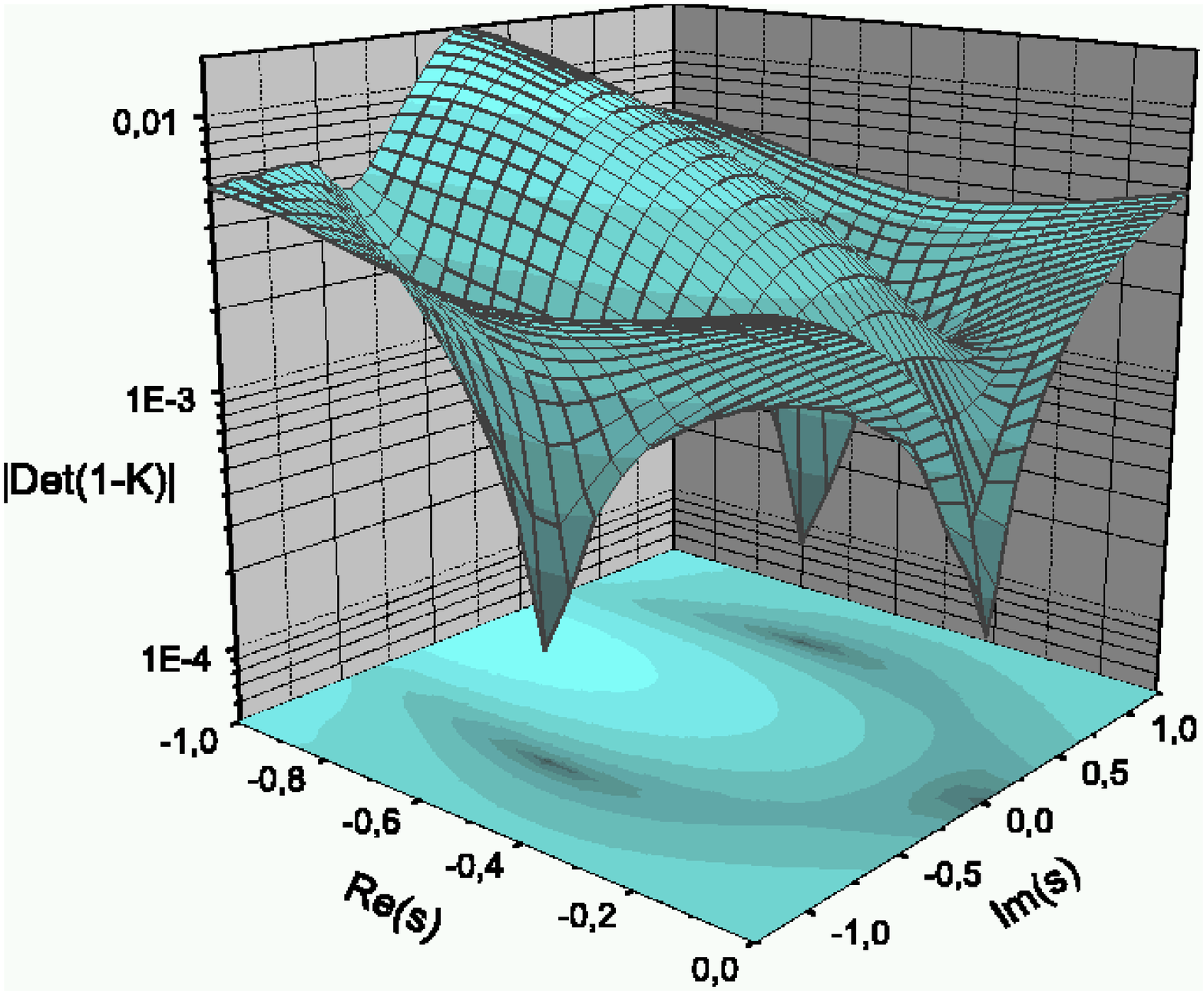,width=7.5cm}\hspace{0.5cm}\epsfig{figure=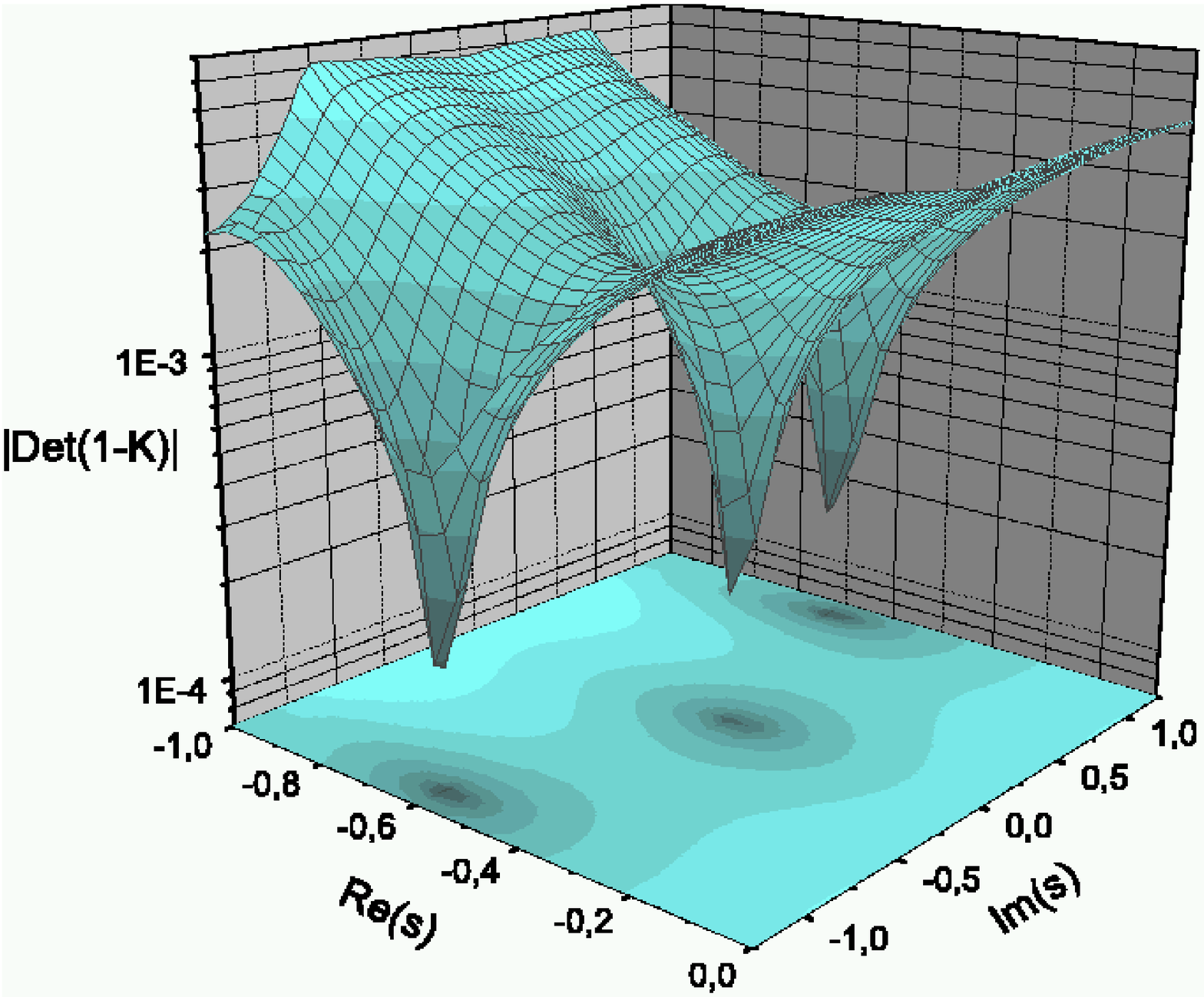,width=7.5cm}}
\caption{\label{fig:sing}Plots of $|Det(\openone-K)|$ for pseudoscalar (left panel) and
scalar (right panel) channels as a function of complex $s$ showing the location of zeroes
in the near timelike region ($\ov{u}u$ case).}
\end{figure}

The approximate rainbow-ladder \BS
meson masses  (discussed above) are not the only singularities in
$G^4$.  Also singularities for general complex values of $s$ show
up.  In order to see the latter it is useful to consider the
numerical solution of Eqs. (\ref{eq:4pt3}). When
discretizing the momenta, these equations become matrix equations
of the form $\G=C+K\G$ and the condition for singularities is that
$Det(\openone-K)=0$, where $K$ is the discretized integral kernel.
A plot of $|Det(\openone-K)|$ as a function of complex $s$ then
shows the positions of such zeroes in each channel.  The near
timelike region studied is $-1GeV^2<\Re(s)<0$,
$-1GeV^2<\Im(s)<1GeV^2$ while the spacelike region has been
studied for $\Re(s)\rightarrow\infty$ and $|\Im(s)|$ increasing
with $\Re(s)$. There are no resonances found in the spacelike
region for any of the channels and also no resonances for the
diquark channels, where $|Det(\openone-K)|$ is largely uniform and
of ${\cal O}(1)$.  Figure \ref{fig:sing} shows the zeroes in the
near timelike region for the pseudoscalar and scalar meson
channels with a $\ov{u}u$ flavor combination.  In the near
timelike region, the only other $\ov{u}u$ flavor resonance is the
vector located on the negative real axis (as reported in the
previous paragraph). The $\ov{s}u$ flavor combination shows a
similar behavior.  It is not immediately clear what the extra
complex conjugate pseudoscalar and scalar resonances are since
they appear to be artifacts of the truncations and approximations
employed. Anyhow, these resonances lie outside the region of
interest for our studies of the light mesons. Nevertheless, the
nature of these `resonances' will have to be explored in future.

\begin{figure}[t]
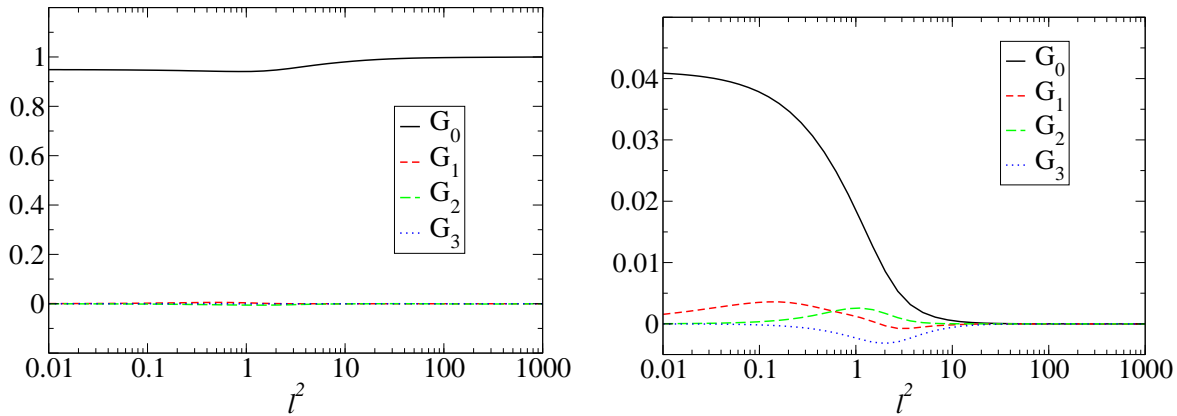

\vspace{1cm}
\mbox{\epsfig{figure=chebr.eps,width=7.5cm}\hspace{0.5cm}\epsfig{figure=chebi.eps,width=7.5cm}}
\caption{\label{fig:cheb}Real and imaginary (left and right panels respectively) parts
of the Chebyshev moments, $G_i^{n,1}(s=1+\imath,t,l^2)$ for real $l^2$.}
\end{figure}

Turning now to the solutions of Eqs. (\ref{eq:4pt3}) we start with
spacelike $l^2$ and show typical results for the $\ov{u}u$ flavor
combination with $s=1+\imath$.  As previously discussed, the
$t$-dependence of the solution is removed by setting the seed term
to $1/\sqrt{2}$ such that the zeroth Chebyshev as
$l^2\rightarrow\infty$ tends to unity ($-3$) for the diquark
(mesonic) components.  We note that for general complex $s$ the
Chebyshev moments of the amplitudes will be complex.  We show both
the real and imaginary parts of the Chebyshev moments of the
diquark $G^{n,1}$ component in Fig.  \ref{fig:cheb}.  Clearly seen
is that the real part of the zeroth Chebyshev dominates primarily
because of the seed terms in Eqs. (\ref{eq:4pt3}). Essentially the 4-pt
function away from resonance positions shows its tree-level form.
Furthermore, the diquark components do not have a resonance in
either the spacelike or near timelike regions and one might
conclude that $G^n$ could be well approximated by its tree-level
term entirely; this will turn out not to be the case.  The mesonic
components for this value of $s$ display the same behavior as
their diquark counterparts.  However, the mesonic sector does have
resonances where the subleading Chebyshev structure and the
deviation from the tree-level form become important.

\begin{figure}[t]
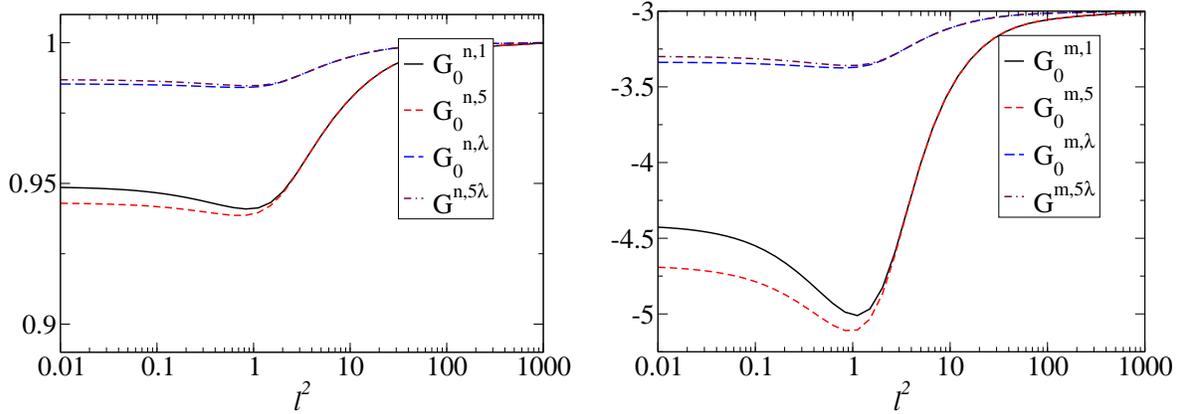

\vspace{1cm}
\mbox{\epsfig{figure=diq.eps,width=7.5cm}\hspace{0.5cm}\epsfig{figure=mes.eps,width=7.5cm}}
\caption{\label{fig:c0}Real parts of the zeroth Chebyshev moment
of the diquark and meson (left and right panels, respectively)
amplitudes $G_0^{(n,m)a}(s=1+\imath,t,l^2)$ displayed as functions
of real $l^2$.}
\end{figure}

We display the real parts of the zeroth Chebyshev amplitudes of
the diquark and meson amplitudes as functions of spacelike $l^2$
for $s=1+\imath$ in Fig. \ref{fig:c0}. This plot specifies the
behavior of these functions away from resonance.  As $l^2$
increases the scalar and pseudoscalar, vector and axialvector
functions converge to the same values.  This is because the
difference between the channels (at this level of approximation)
is the sign of the factor $s_1s_2$ in the kernel, which as
$l^2\rightarrow\infty$ rapidly vanishes.  A comparison of the
diquark and mesonic channels reveals that the two have similar
shaped curves, though with different amplitudes. This can be
attributed to the similarity in the kernels of the two channels.

\begin{figure}[t]
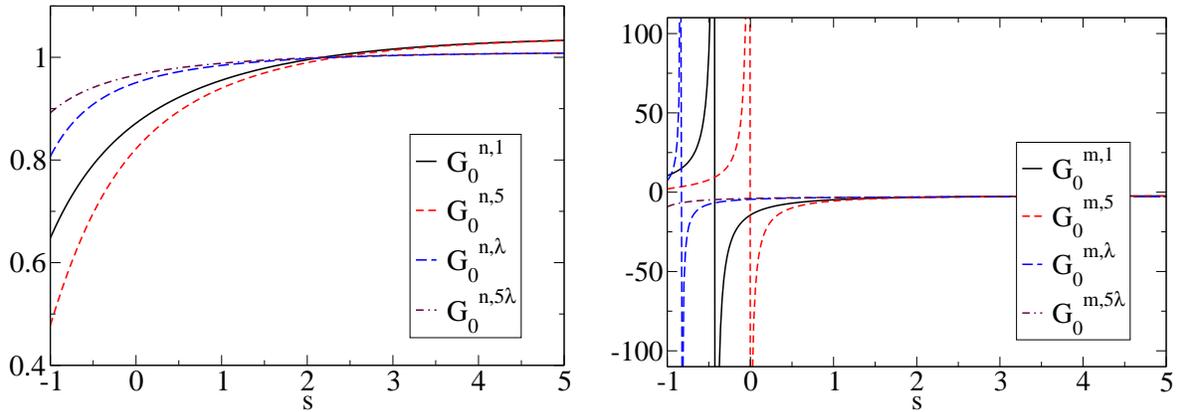

\vspace{1cm}
\mbox{\epsfig{figure=diqs.eps,width=7.5cm}\hspace{0.5cm}\epsfig{figure=mess.eps,width=7.5cm}}
\caption{\label{fig:sings}Zeroth Chebyshev moments of the diquark
and meson (left and right panels, respectively) amplitudes
$G_0^{(n,m)a}(s,t,l^2=0)$ plotted as functions of real $s$.}
\end{figure}

It is, furthermore, instructive to plot $G_0^{(n,m)a}(s,t,l^2=0)$ as
a function of real $s$ to highlight the resonance structure (cf.
Fig. \ref{fig:sings}).  The diquark correlations show no resonance
structure, but crucially do show significant variation with $s$,
implying that the tree-level approximation is not reliable for low
$s$. The mesonic correlations show resonance structures as
expected. It is tempting to approximate these curves by a simple
$1/(s+m^2)$ shape, but closer inspection reveals that there they
are indeed more complicated.  This is an important point to note
since the deviations from the pole term are significant.

\begin{figure}[t]
\vspace{1cm}
\mbox{\epsfig{figure=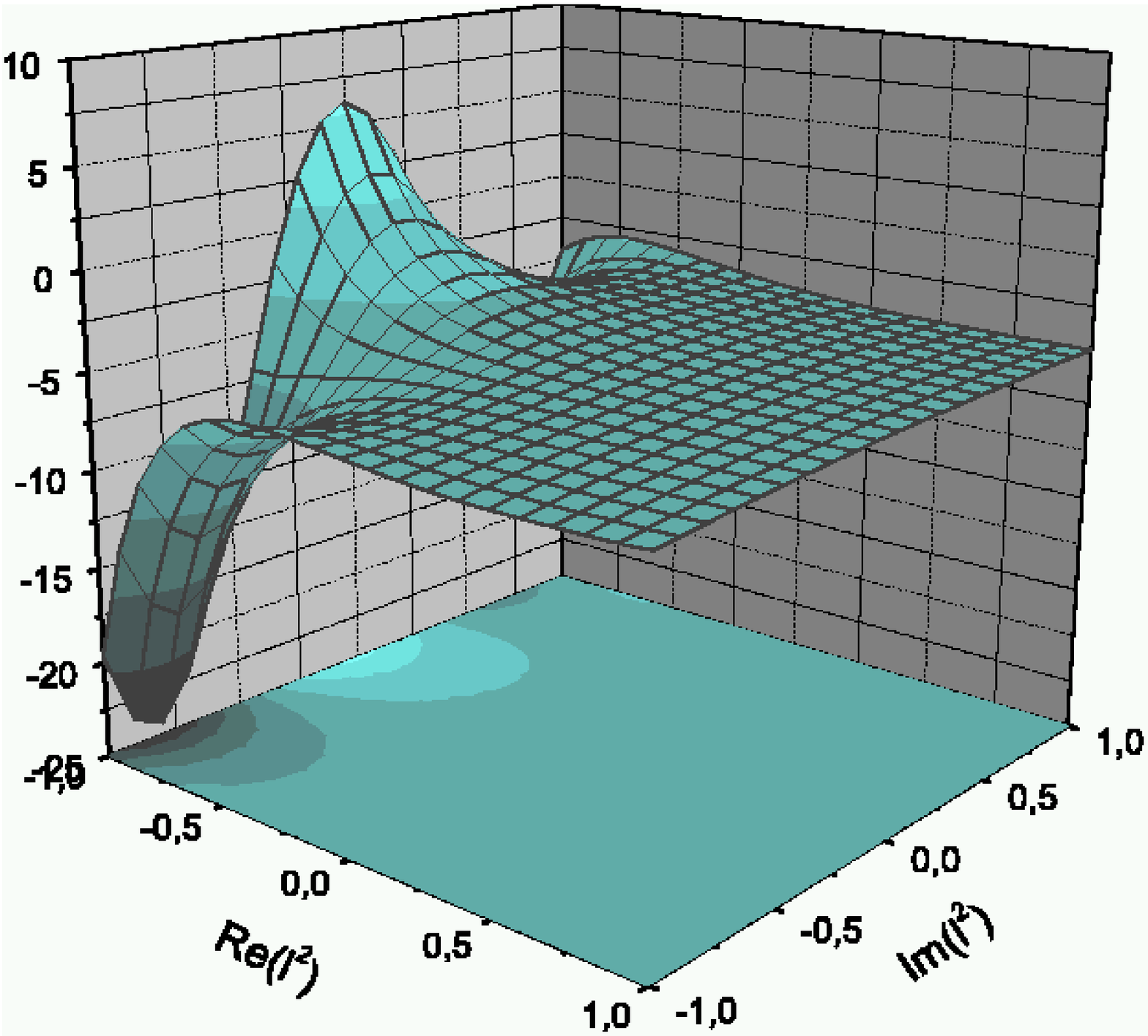,width=7.5cm}\hspace{0.5cm}\epsfig{figure=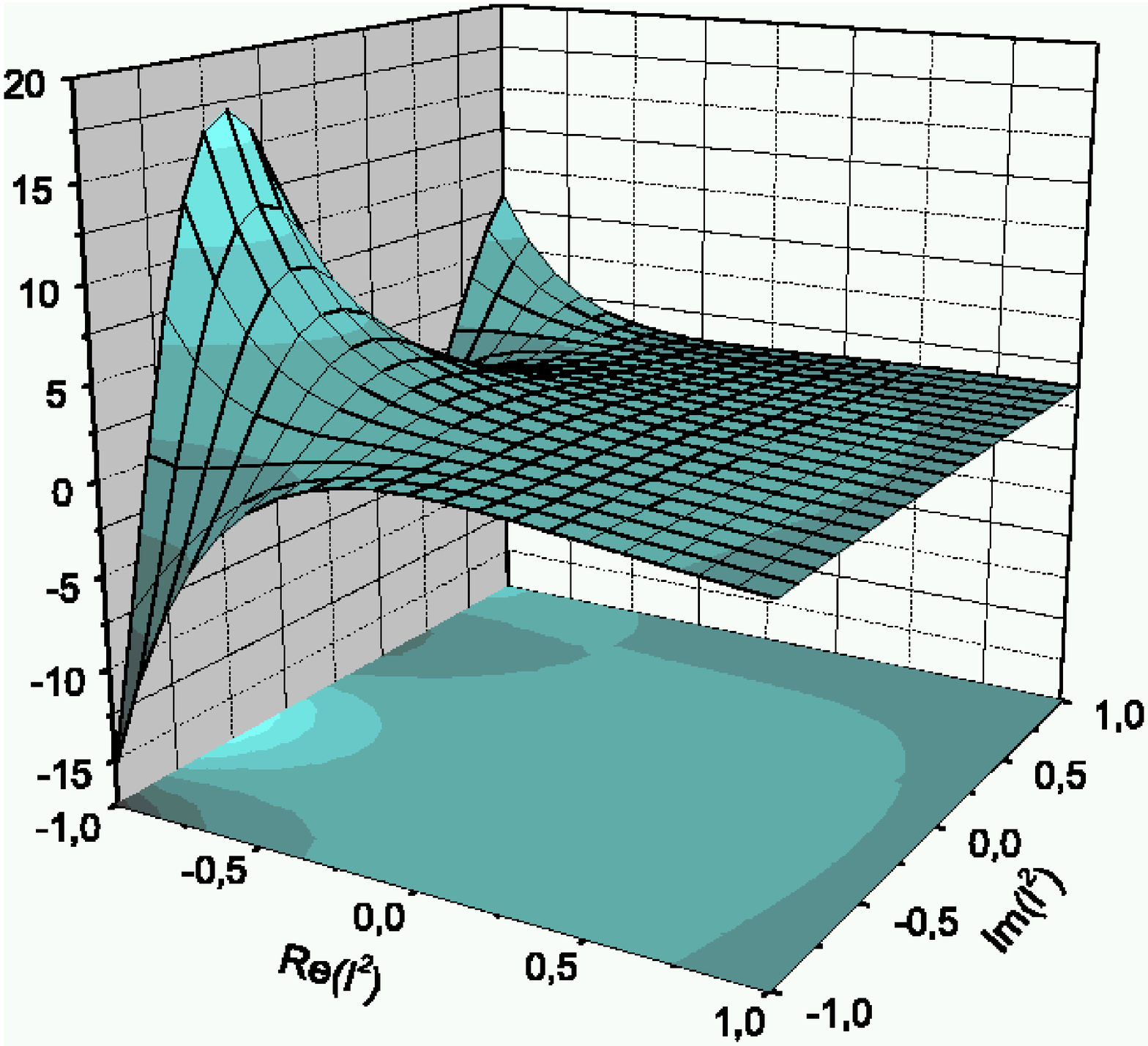,width=7.5cm}}
\caption{\label{fig:cmplx}Real and imaginary (left and right
panels, respectively) parts of the zeroth Chebyshev moments of the
$\pi$ amplitude $G_0^{m,5}(s=1+\imath,t,l^2)$ plotted as functions
of complex $l^2$ in the vicinity of $l^2=0$.}
\end{figure}

Finally, we show the functions $G_i^{(n,m)a}(s,t,l^2)$ for the case
of complex $l^2$. Recall that in the kernel we need the amplitude
for $l^2=(k+q/2\pm P/4)^2$, where $k,q$ are integration variables,
$P^2=-M^2$, while $M$ is the complex mass of the meson under
consideration.  The factor $P/4$ does, however, mean that the
deviation from the spacelike axis will only be slight. A
consideration of the exponential interaction shows that the
functions will be varying most significantly in the timelike and
imaginary directions. Thus it is important that these deviations
are taken into account, if only to assess their relevance to the
final results. We show $G_0^{m,5}(s=1+\imath,t,l^2)$ (the leading
Chebyshev moment of the $\pi$) for complex $l^2$ in the vicinity of
zero in Fig. \ref{fig:cmplx}.  It is seen is that the functions
vary smoothly and -- as expected -- the variation becomes larger
when progressing into the complex region.

\section{Summary and Conclusions}
In this work, a proposed approximate form for the connected four-point 
quark-antiquark Green's function ($G^4$) has been studied.
The motivation for the study is an extension of the conventional
\DS-- \BS scheme to explore unquenching effects in the light meson sector.  
The expression for the approximated four-point
quark-antiquark Green's function has been derived and shown
to be consistent with asymptotic limits.  Furthermore, it
reproduces the leading structure of the ladder \BS equation for
the (lightest) pseudoscalar mesons.  The resulting equations 
have been solved numerically and typical results presented for 
both real and complex momenta.

Most of the approximations employed in this work have been made 
for heuristic purposes and should be investigated in more detail 
in subsequent work.  Nevertheless, our calculations demonstrate 
that the scheme proposed already gives acceptable results for the light meson
sector and excludes resonances in the diquark channel.  Clearly 
these initial results provide a basis for the study of 
dynamical meson decay and possible multiquark structures.

\section*{Acknowledgments}
\noindent
The authors thank R. Alkofer for a critical reading of
the manuscript and valuable suggestions. This work has been performed
under grant no. COSY 41139452.


\begin{thebibliography}{99}

\bibitem{lattice1}

M.~Golterman, and Y.~Shamir, [arXiv:hep-lat/0404001]; 
J.~Smit, Nucl. Phys. Proc. Suppl. {\bf 17} (1990) 3; 
D.~N.~Petcher, [arXiv:hep-lat/9301015]; 
M.~Golterman, Nucl. Phys. Proc. Suppl. {\bf 94} (2001) 189, [arXiv:hep-lat/0011027]; 
J.~W.~Negele \textit{et al.}, Nucl. Phys. Proc. Suppl. {\bf 128} (2004) 170, [arXiv:hep-lat/0404005]; 
K.~Schilling, H.~Neff, and T.~Lippert, [arXiv:hep-lat/0401005].


\bibitem{maris03}

P.~Maris, and C.~D.~Roberts, Int. J. Mod. Phys. {\bf E12} (2003) 297, [arXiv:nucl-th/0301049].


\bibitem{roberts00}

C.~D.~Roberts and S.~M.~Schmidt, Prog.\ Part.\ Nucl.\ Phys.\ {\bf 45} (2000) S1, [arXiv:nucl-th/0005064].


\bibitem{alkofer00}

R.~Alkofer, and L.~v.~Smekal, Phys. Rep. {\bf 353} (2001) 281, [arXiv:hep-ph/0007355].


\bibitem{roberts94}

C.~D.~Roberts, and A.~G.~Williams, Prog. Part. Nucl. Phys. {\bf 33} (1994) 477, [arXiv:hep-ph/9403224].


\bibitem{munczek83}

H.J.~Munczek, and A.M.~Nemirovsky, Phys. Rev. {\bf D28} (1983) 181.

\bibitem{jain93}

P.~Jain, and H.~J.~Munczek, Phys. Rev. {\bf D48} (1993) 5403; 
H.~J.~Munczek, and P.~Jain, Phys. Rev. {\bf D46} (1991) 438;
P.~Jain, and H.~J.~Munczek, Phys. Rev. {\bf D44} (1991) 1873.

\bibitem{burden96}

C.~J.~Burden, L.~Qian, C.~D.~Roberts, P.~C.~Tandy, M.~J.~Thomson, Phys. Rev. {\bf C55} (1997) 2649, [arXiv:nucl-th/9605027].


\bibitem{maris97b}

P.~Maris, C.~D.~Roberts, and P.~C.~Tandy, Phys. Lett. {\bf B420} (1998) 267, [arXiv:nucl-th/9707003].


\bibitem{delbourgo79}

R.~Delbourgo, and M.~D.~Scadron, J. Phys. {\bf G5} (1979), 1621.

\bibitem{maris99a}

P.~Maris, and P.~C.~Tandy, Phys. Rev. {\bf C60} (1999) 055214, [arXiv:nucl-th/9905056].


\bibitem{me02}

R.~Alkofer, P.~Watson, and H.~Weigel, Phys. Rev. {\bf D65} (2002) 094026, [arXiv:hep-ph/0202053].


\bibitem{maris02}

P.~Maris, and P.~C.~Tandy, Phys. Rev. {\bf C65} (2002) 045211, [arXiv:nucl-th/0201017].


\bibitem{maris99b}

P.~Maris, and P.~C.~Tandy, Phys. Rev. {\bf C61} (2000) 045202, [arXiv:nucl-th/9910033].


\bibitem{bhagwat03}

M.~S.~Bhagwat, M.~A.~Pichowsky, C.~D.~Roberts, and P.~C.~Tandy, Phys. Rev. {\bf C68} (2003) 015203, [arXiv:nucl-th/0304003].


\bibitem{bender96}

A.~Bender, C.~D.~Roberts, and L.~v.~Smekal, Phys. Lett. {\bf B380} (1996) 7, [arXiv:nucl-th/9602012].


\bibitem{detmold02}

A.~Bender, W.~Detmold, C.~D.~Roberts, and A.~W.~Thomas, Phys. Rev. {\bf C65} (2002) 065203, [arXiv:nucl-th/0202082].


\bibitem{ongoing04}

P.~Watson, W.~Cassing, and P.~C.~Tandy, to be published.

\bibitem{jarecke02}

D.~Jarecke, P.~Maris, and P.~C.~Tandy, Phys. Rev. {\bf C67} (2003) 035202, [arXiv:nucl-th/0208019].


\bibitem{pennington03}

M.~R.~Pennington, [arXiv:hep-ph/0309228].


\bibitem{nakanishi69}

N.~Nakanishi, Supp. Prog. Theo. Phys. {\bf 43} (1969) 1.

\bibitem{kekez00}

D.~Kekez, D.~Klabucar, and M.~D.~Scadron, J. Phys. {\bf G26} (2000) 1335, [arXiv:hep-ph/0003234].


\bibitem{kogut74}

J.~Kogut, and L.~Susskind, Phys. Rev. {\bf D10} (1974) 3468.

\bibitem{scadron03}

M.~D.~Scadron, G.~Rupp, F.~Kleefeld, and E.~v.~Beveren, Phys. Rev. {\bf D69} (2004) 014010; 
erratum-ibid. {\bf D69} (2004) 059901, [arXiv:hep-ph/0309109].


\bibitem{tornqvist02}

N.~A.~T\"ornqvist, "The lightest scalar nonet", in Proc. "Chiral Fluctuations in Hadronic Matter" workshop, Orsay, 16-20 Sept. 2001, p267, edited Z.~Aouissat et al., [arXiv:hep-ph/0201171].


\bibitem{jaffe77}

R.~L.~Jaffe, Phys. Rev. {\bf D15} (1977) 267.

\bibitem{black00}

D.~Black, A.~H.~Fariborz, S.~Moussa, S.~Nasri, and J.~Schechter, Phys. Rev. {\bf D64} (2001) 014031, [arXiv:hep-ph/0012278].


\bibitem{close02}

F.~E.~Close, and N.~A.~T\"ornqvist, J. Phys. {\bf G28} (2002) R249, [arXiv:hep-ph/0204205].


\bibitem{tandy97}

P.C.~Tandy, Prog. Part. Nucl. Phys. {\bf 39} (1997) 117, [arXiv:nucl-th/9705018].


\bibitem{bhagwat02}

M.~S.~Bhagwat, M.~A.~Pichowsky, and P.~C.~Tandy, [arXiv:hep-ph/0212276].


\bibitem{cvitanovic76}

P.~Cvitanovic, Phys. Rev. {\bf D14} (1976) 1536.

\bibitem{cahill89}

R.~T.~Cahill, J.~Praschifka, and C.~J.~Burden, Aust. J. Phys. {\bf 42} (1989) 161.

\end{thebibliography}
\end{document}